\documentclass[]{aa}  
\usepackage{graphicx}
\usepackage{txfonts}
\usepackage{natbib}
\usepackage{color}  
\usepackage{ulem}   

\begin{document}

   \title{CME liftoff with high-frequency fragmented type II burst emission}


   \author{S.Pohjolainen\inst{1}
          \and
          J. Pomoell\inst{2} 
          \and 
          R. Vainio\inst{2} 
          }

   \offprints{S. Pohjolainen}

   \institute{Department of Physics and Astronomy, University of Turku,
              Tuorla Observatory, 21500 Piikki\"o, Finland\\
              \email{silpoh@utu.fi}
              \and
              Department of Physics, PO Box 64,
              00014 University of Helsinki, Finland\\
              \email{jens.pomoell@helsinki.fi, rami.vainio@helsinki.fi}
             } 

   \date{Accepted August 2008}

  \abstract{}{}{}{}{} 
 
  \abstract
    {} 
   {Solar radio type II bursts are rarely seen at frequencies higher than 
    a few hundred MHz. Since metric type II bursts are thought to be
    signatures of propagating shock waves, it is of interest to know how 
    these shocks, and the type II bursts, are formed. In particular, how 
    are high-frequency, fragmented type II bursts created? Are there 
    differences in shock acceleration or in the surrounding medium that 
    could explain the differences to the ``typical'' metric type IIs?}
    {We analyse one unusual metric type II event in detail, with comparison
     to white-light, EUV, and X-ray observations. As the radio event 
     was associated with a flare and a coronal mass ejection (CME), we
     investigate their connection. We then utilize numerical MHD simulations 
     to study the shock structure induced by an erupting CME in a model 
     corona including dense loops.}
    {Our simulations show that the fragmented part of the type II 
    burst can be formed when a coronal shock driven by a mass ejection 
    passes through a system of dense loops overlying the active region.
    To produce fragmented emission, the conditions for plasma emission
    have to be more favourable inside the loop than in the interloop area. 
    The obvious hypothesis, consistent with our simulation model, is 
    that the shock strength decreases significantly in the space between 
    the denser loops. The later, more typical type II burst appears when 
    the shock exits the dense loop system and finally, outside the active 
    region, the type II burst dies out when the changing geometry no longer
    favours the electron shock-acceleration.
     }
    {}
   
   \maketitle
%

\section{Introduction}

Radio type II bursts are observed in association with flares and
coronal mass ejections (CMEs). These bursts can be observed at
metric wavelengths (coronal type II bursts) and at decameter and 
longer wavelengths (interplanetary type II bursts). The mechanism 
behind the bursts is generally assumed to be a propagating shock 
which creates electron beams that excite Langmuir waves, which 
in turn convert into radio waves at the local plasma frequency 
and its second harmonic \citep{melrose80,cairns03}.
As shocks can be formed in various ways (for an overview see, e.g.,
Warmuth, \citeyear{warmuth07}; for the terminology see, e.g., 
Vr\u{s}nak, \citeyear{vrsnak05}), it is not evident that all 
solar radio type II bursts are formed in  the same way. 

Most of the interplanetary type II bursts are thought to be created 
by CME-driven shocks, but there is observational evidence that 
at least some coronal metric type II bursts are ignited by 
smaller-scale processes associated with the flare energy release, 
such as high-speed plasma jets/blobs or loop ejections/expansions 
\citep{klein99,pohjolainen01,khan02,klassen03,dauphin06,pohjolainen08}.
Statistically, there is support for the idea that metric type II
bursts have their root cause in fast coronal mass ejections 
\citep{cliver99}, but also that they are not caused by shocks 
driven in front of CMEs \citep{cane05}.         
 
Metric type II bursts can be observed in dynamic radio spectra 
as slowly (0.1\,--\,1.0 MHz s$^{-1}$) drifting emission lanes 
\citep{nelson85}). The start frequency of metric type II bursts 
is usually at about 100\,--\,200 MHz, and the bursts have a typical 
duration of 2\,--\,18 min \citep{subra06}.  
In some cases it is possible to separate a slowly drifting ``backbone'' 
in the burst emission, with fast-drifting ($\sim$ 10 MHz s$^{-1}$) 
emission stripes shooting up and down from the backbone 
\citep {mann05}.
These features have been named as ``herringbones'', and they bear
some resemblance to radio type III bursts which are signatures 
of outflowing electron beams. However, the drift rates of type III
bursts have been found to be higher than those of herringbones
\citep{mann02}. 

In this paper, we analyse in detail one metric type II burst that 
occurred on 13 May 2001. The burst started at an unusually high 
frequency, proceeded showing fragmented and curved emission 
bands, which were visible at the fundamental and second harmonic
plasma frequencies. As the radio burst was associated with a flare 
and a CME, we do a multi-wavelength analysis in order to identify 
the origin of the shock that formed the radio type II burst.
We then use numerical MHD simulations to verify the results
from the data analysis.


   \begin{figure*}
   \includegraphics[width=9.5cm,angle=0]{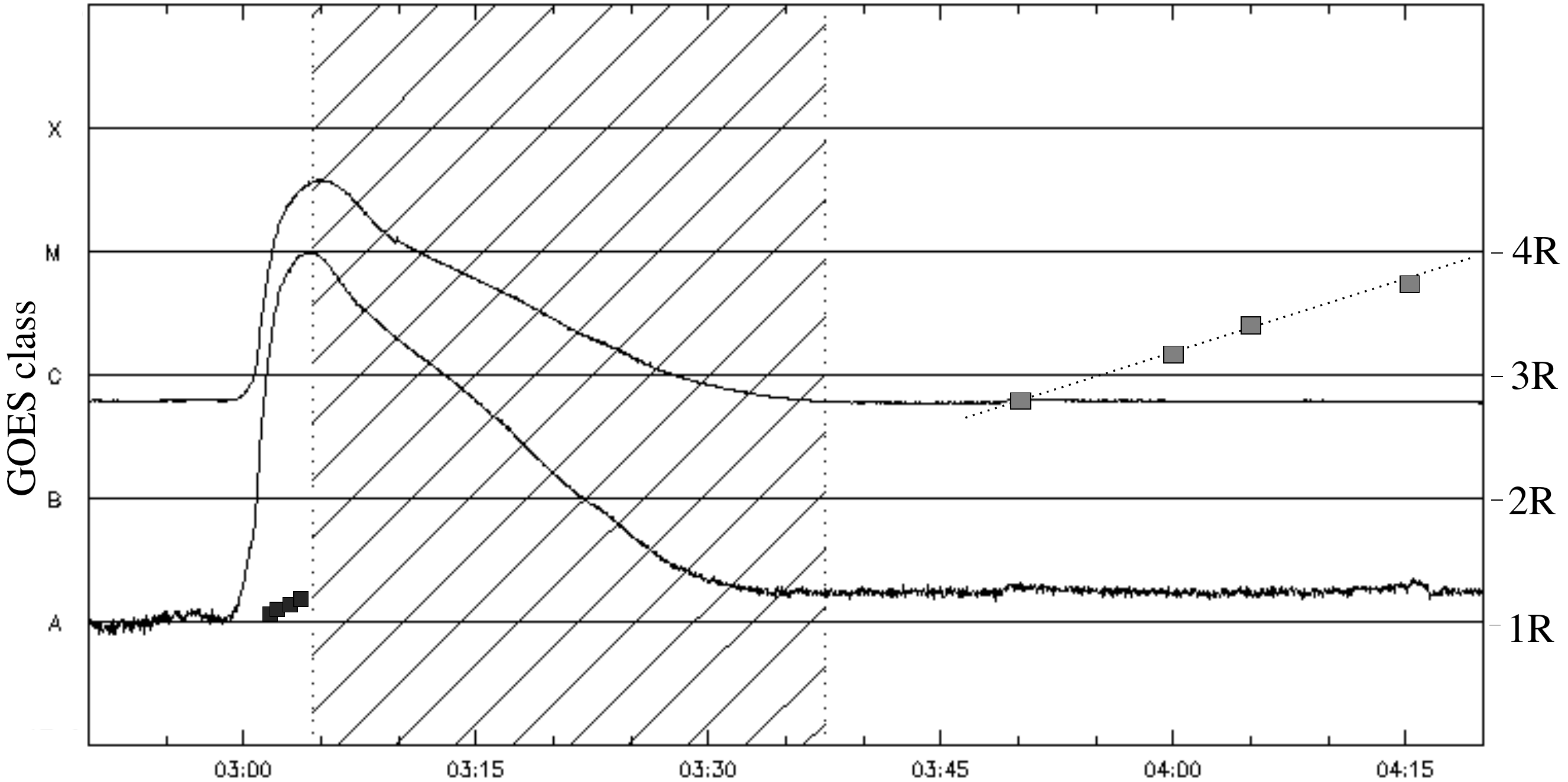}
   \includegraphics[width=9cm,angle=0]{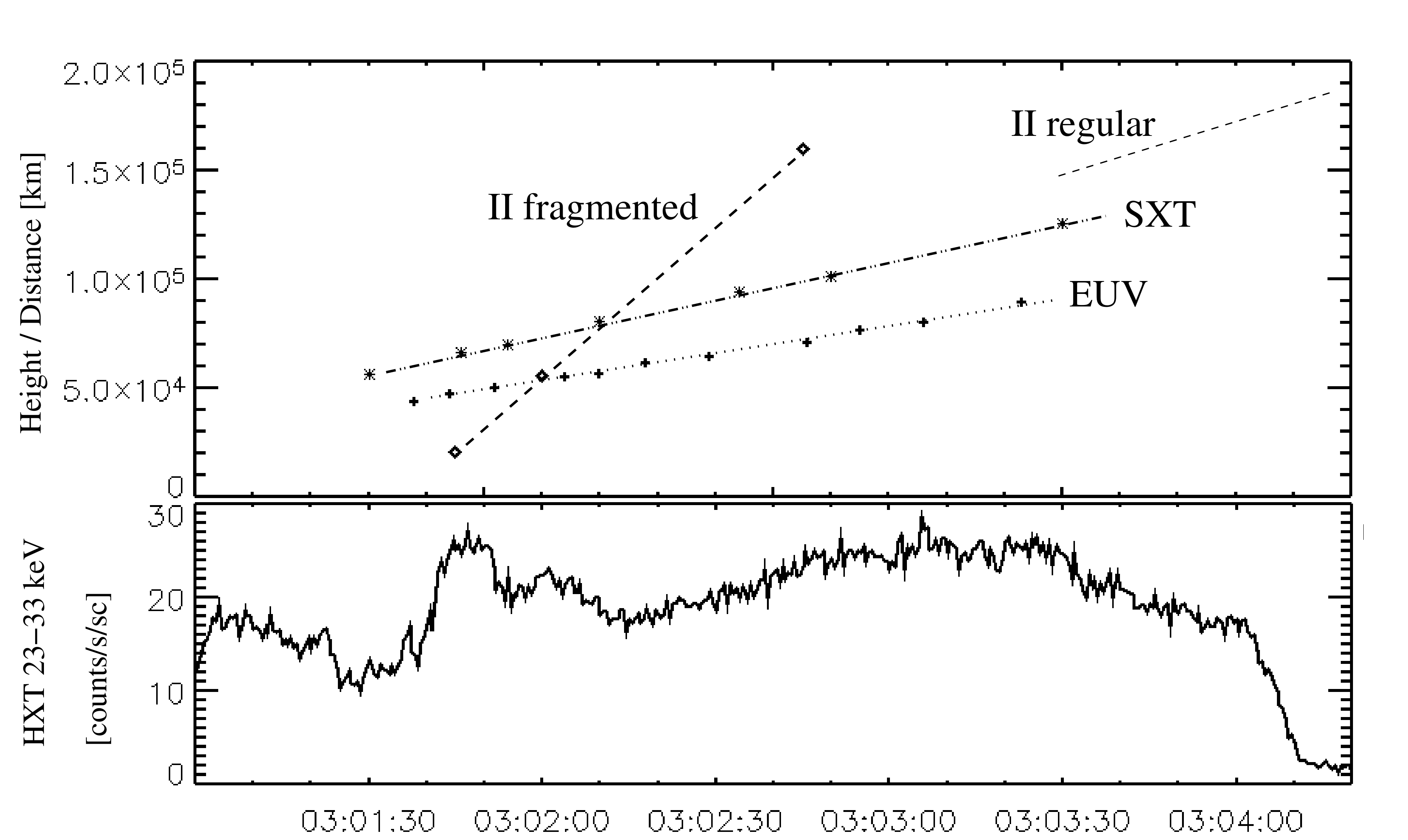}
      \caption{Left: GOES soft X-ray flux curve at 02:50\,--\,04:20 UT, 
               with flare class classification A--X. The dashed region 
               indicates Yohkoh satellite night time, when no 
               observations are available from the experiment.
               Plane-of-the-sky heliocentric CME front distances (boxes 
               combined with a dotted line) are from the LASCO CME Catalog.
               Black boxes between 03:01:30 and 03:03:30 UT show the 
               projected distances of the soft X-ray loop-like front,   
               measured from the center of the eruption region at 
               1 R$_{\odot}$.
               Right: Detailed plot of the estimated distances/heights 
               measured in the time interval 03:01:00\,--\,03:04:20 UT. 
               Projected distances (km) from the eruption center are shown 
               for the SXT loop-like front (stars/dash-dotted line) and the 
               EUV 'blob' (crosses/dotted line). The estimated type II 
               burst heights are calculated with the 10$\times$Saito 
               density model for the fragmented part (the three height--time 
               data points, marked with diamonds, are from the start times 
               of each three fragment. The type II heights for the later 
               regular part (dashed lines) are calculated with the hybrid 
               density model (corresponding to 5$\times$Saito) which 
               describes a ``standard'' solar atmosphere. Applying a 
               high-density model like 10$\times$Saito for the 
               regular type II part at $\lesssim$ 130 MHz would imply  
               at least streamer densities, see text for details.
               Yohkoh HXT hard X-ray counts in the 23\,--\,33 keV 
               energy channel are also shown (bottom panel). Note that 
               the drop in counts near 03:04 UT is due to the satellite 
               entering night time.
              }
         \label{fig1}
   \end{figure*}

\section{Observations}

A GOES M3.6 class flare was observed to start at 02:58 UT on 13 May 
2001 in NOAA AR 9455 (S18\,W01). The flare maximum occurred at 
03:04 UT, and at 03:50 UT a coronal mass ejection was detected in  
the LASCO \citep{brueckner95} C2 images. The CME front moved toward 
the South at a speed of 430 km s$^{-1}$ (LASCO CME Catalog, both the 
linear and second-order fits to the height--time data give similar 
speeds). Backward-extrapolation of the heights indicates a CME start 
time near 03:00 UT. Fig.~\ref{fig1} shows the GOES soft X-ray flux 
curve, with the CME plane-of-the-sky heights during 02:50\,--\,04:20 UT.
Yohkoh HXT \citep{kosugi91} hard X-ray counts are shown at 
03:01:00\,--\,03:04:20 UT, before satellite night ended the 
observations.

\begin{figure}[]
   \centering
   \includegraphics[width=8cm]{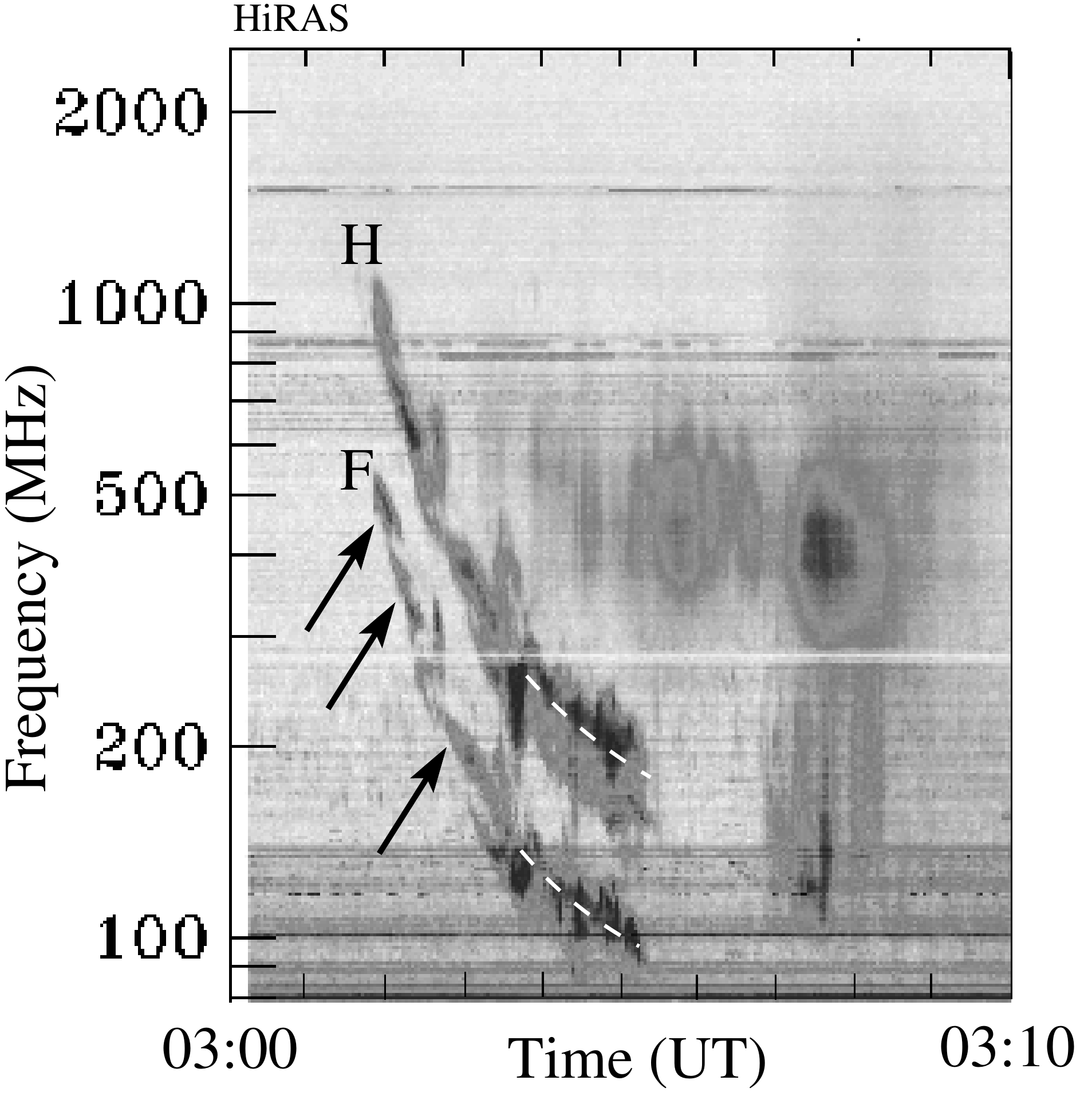}
      \caption{HiRAS dynamic spectrum from the 25-2500 MHz frequency
               range, at 03:00--03:10 UT on 13 May 2001. 'F' notes
               emission at the fundamental and 'H' at the second
               harmonic plasma frequency. Arrows point to the 
               fragmented emission bands and dashed white lines outline 
               the later-appearing ``regular'' type II burst lanes.     
              }
         \label{fig2}
   \end{figure}

Radio emission at decimetric--metric wavelengths was observed to start 
at 03:01:45 UT. It consisted of a type II burst that was clearly visible 
as two emission lanes, starting near 500 and 1000 MHz. It was followed 
by bursty continuum emission at 700\,--\,300 MHz. These features are shown 
in the HiRAS dynamic spectrum that covers a wide, 2500--25 MHz frequency 
range, in Fig.~\ref{fig2}. The type II burst emission ended abruptly at 
03:05:15 UT near 90 MHz at the fundamental emission band (visible also 
in the RSTN observations at 180--25 MHz). No interplanetary type 
II emission was observed at lower frequencies, as verified from the 
Wind WAVES \citep{bougeret95} observations at 14 MHz--20 kHz.

  \subsection{Fragmented radio type II burst}

The radio type II burst showed a pair of fundamental and second 
harmonic emissions, which clearly marks it as plasma emission,
\begin{equation}
f_p = 9000 \sqrt{n_e}, 
\end{equation}
where the local electron plasma frequency $f_p$ is expressed in Hz 
and the electron density $n_e$ in cm$^{-3}$. As plasma emission at the 
fundamental is directly related to the local electron density, 
we can use the spectral observations to calculate electron densities 
and to estimate burst source heights and speeds. Height estimates 
require, however, the use of atmospheric density models, which are 
not unambiguous, see e.g., \citet{vrsnak04a} and \citet{pohjolainen07}.

The 500 MHz start frequency (fundamental plasma emission) is unusually 
high for a type II burst \citep{lin06}. During the first two minutes of 
the burst, the emission was fragmented into separate, curved 
emission bands. The frequency drifts (fundamental emission) 
within the fragments were between 1.8 and 4.3 MHz s$^{-1}$. 
These drift rates are high compared to type II bursts in general, 
but consistent with drift rates found for high-frequency type II
bursts, see Fig. 2a in \citet{vrsnak02}. 
The first ``fragment'' showed emission between 500 and 420 MHz, which 
correspond to densities in the range of  
\mbox{2\,--\,3 $\times$ 10$^9$ cm$^{-3}$}. The second fragment near 
03:02 UT gives densities \mbox{1\,--\,2 $\times$ 10$^9$ cm$^{-3}$} 
(400\,--\,310 MHz), and the third near 03:03 UT gives densities  
\mbox{2\,--\,6 $\times$ 10$^8$ cm$^{-3}$} (220\,--\,130 MHz). 
These values indicate that the source regions were dense, similar 
to active region loops. Rough estimates for heliocentric heights, 
using atmospheric density models up to ten-times Saito \citep{saito70}, 
give values in the range of 1.03\,--\,1.43 R$_{\odot}$. These estimates
are suggestive, since the models do not give heights for erupting 
structures. 
The calculated type II burst source heights (10$\times$Saito densities) 
are shown at the start times of the three fragmented emission 
bands in Fig.~\ref{fig1}. The burst velocity, using these heights, is
as high as 2\,300 km s$^{-1}$, which probably reflects the gradual 
density gradient in the Saito density model. For comparison, if we 
calculate the burst speed from a steeper density gradient, using the 
emission frequency (500 MHz) and the general drift-rate of the bands 
(2.4 MHz s$^{-1}$), and the scale height of 73\,000 km (hydrostatic 
isothermal scale height at 1.25 MK coronal temperature and 
heliocentric height of 1.1  R$_{\odot}$), we get a burst speed of 
$\approx$ 1\,000 km s$^{-1}$ (for the method see, e.g., 
Pohjolainen et al., \citeyear{pohjolainen07}).

At 03:03:35 UT the type II lanes got wider and started to 
look like a typical type II burst. The start frequency of this 
``regular'' part was 130 MHz at the fundamental, and the frequency drift 
was about 0.4 MHz s$^{-1}$. This is within the usual drift rates 
observed for type II bursts (between 0.1 and 1.0 MHz s$^{-1}$). 
To estimate the height for this part of the type II burst we use 
the hybrid density model by \citet{vrsnak04a}. 
This model gives density values similar to five-times Saito in 
the low corona but also links coronal densities to those in the 
interplanetary space. 
The heliocentric heights for the burst source 
are 1.22 R$_{\odot}$ at the beginning of the burst (130 MHz), 
and 1.34 R$_{\odot}$ at the end of the burst at 03:05:15 UT (90 MHz).
A height--time trajectory for this regular type II burst part is 
also shown in Fig.~\ref{fig1}. Applying a high-density model like 
10$\times$Saito also for the regular type II part would make the
two type II burst tracks converge, but this would imply that the
regular part was formed at least at streamer densities,  
above 1.4 R$_{\odot}$.
If the burst source was moving along the density gradient the deduced
speed from the hybrid model is approximately 840 km s$^{-1}$. If the 
direction of motion differed from that, the speed could have been 
higher (an angle of 45$^{\circ}$ between the driver velocity vector 
and the density gradient produces a speed 1.4 times higher).   

   \begin{figure}[]
   \centering
   \includegraphics[width=8cm]{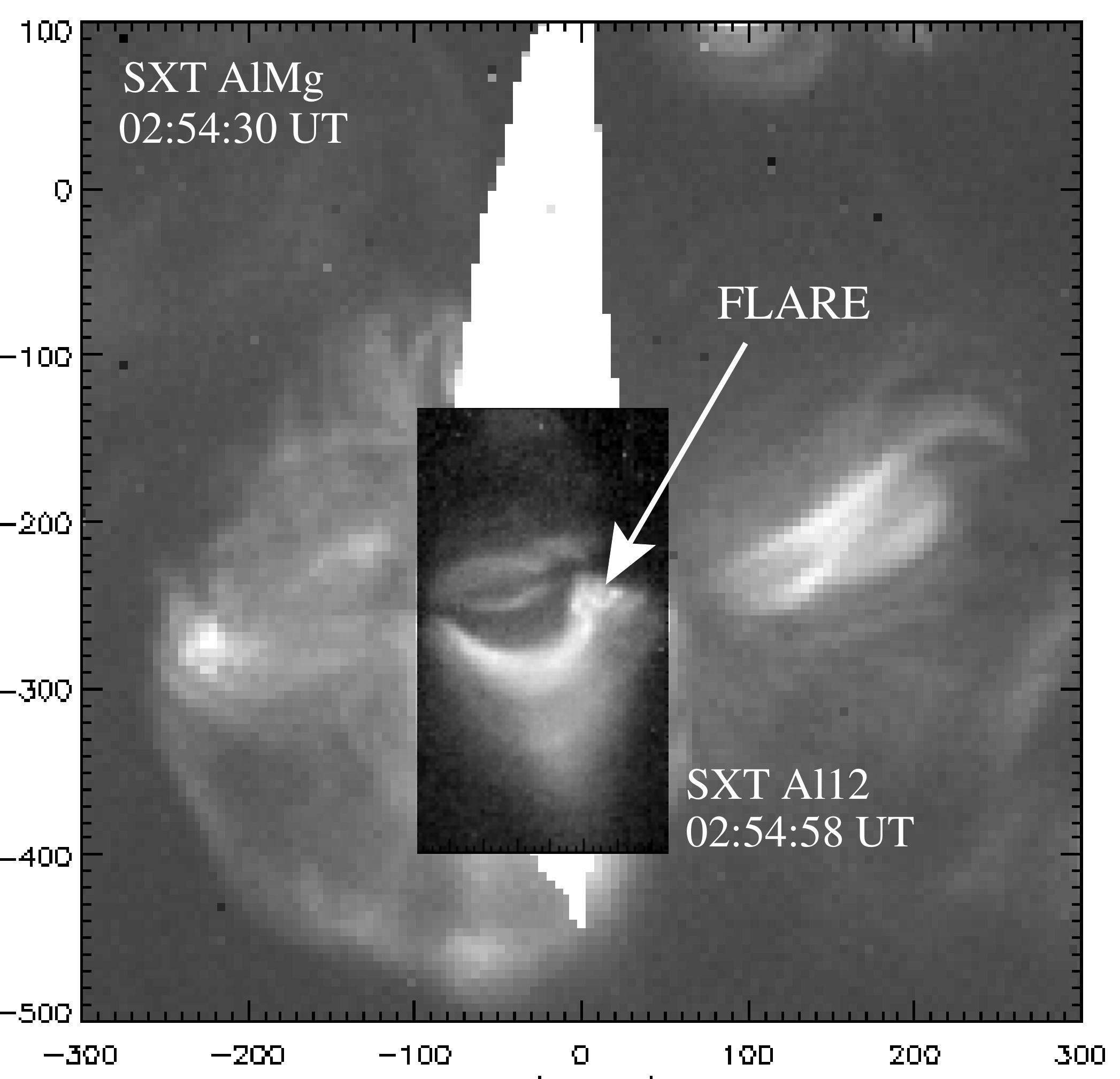}
      \caption{Yohkoh SXT composite of images taken with two different 
       filters and field of views at 02:54 UT on 13 May 2001.    
       It shows the pre-eruption flare region and interconnecting 
       loops. Axes are in arc seconds and white areas
       in the image mark saturated pixels.
   }
   \label{fig3}
   \end{figure}
%

   \begin{figure}[]
   \centering
   \includegraphics[width=6cm]{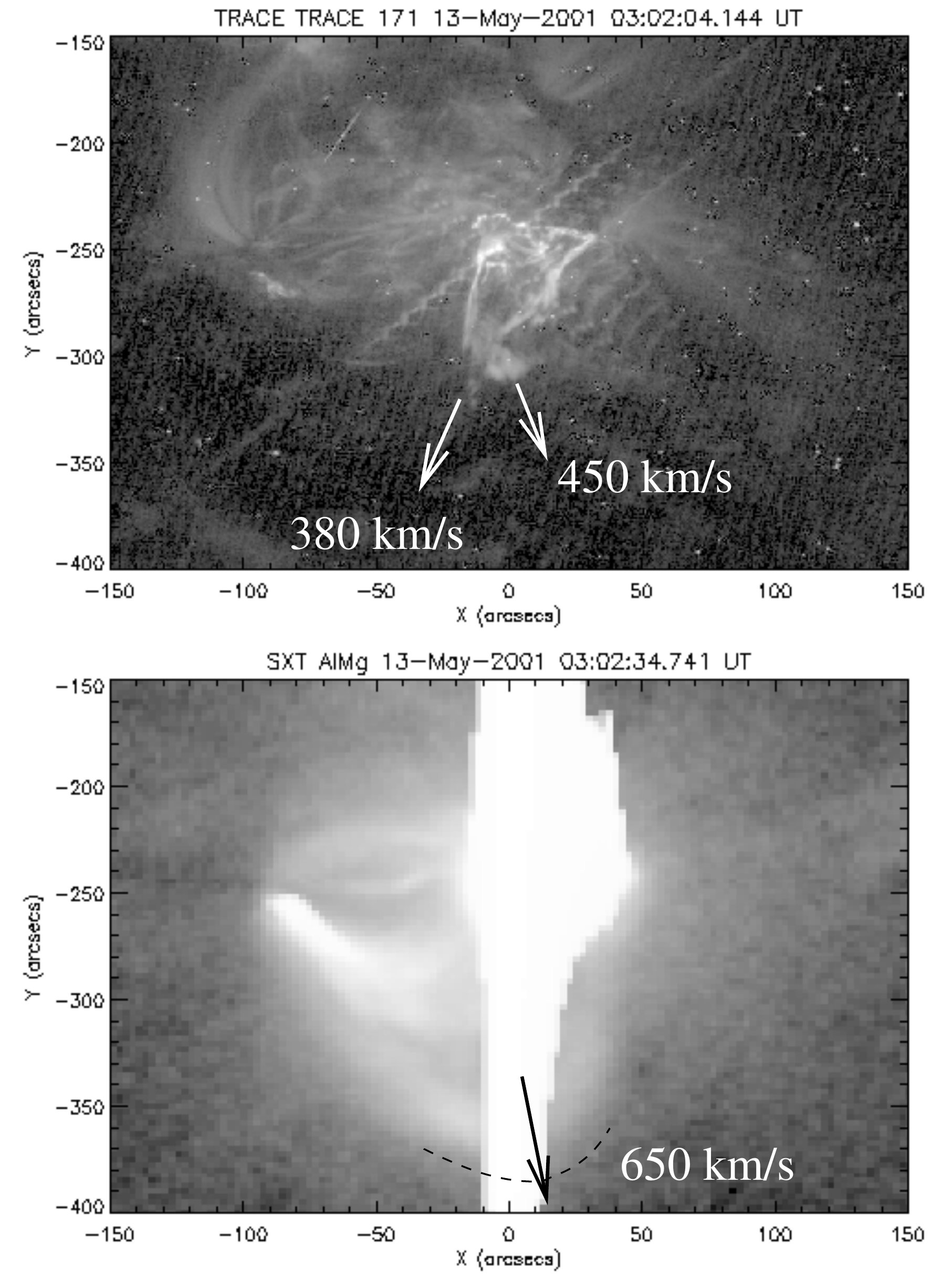}
      \caption{TRACE EUV image at 03:02:04 UT (top), showing the erupting 
              filament. Most of the filament material moved toward the 
              Southeast (speed $\approx$ 380 km s$^{-1}$), but a blob-like 
              structure moved more to the Southwest (speed $\approx$ 
              450 km s$^{-1}$). 
              Arrows give the direction of motion for these structures. 
              Yohkoh SXT image (bottom) shows the same region in
              soft X-rays. The pointing for TRACE seems to be off 
              from that of Yohkoh and so point-to point comparison
              is ambiguous.
              }
         \label{fig4}
   \end{figure}
%

\subsection{Eruptive flare}

The pre-eruption Yohkoh SXT \citep{tsuneta} images show several 
loop systems, connected both to the active region and regions nearby 
(Fig.~\ref{fig3}). Small-scale pre-flare brightenings were observed before 
the flare start at 02:58 UT. The flare region is indicated with an arrow 
in Fig.~\ref{fig3}. 

EUV images taken with TRACE \citep{handy} show a filament eruption, in
which most of the material is moving toward the Southeast. The 
outermost front of the filament moves with a projected speed of 
about 380 km s$^{-1}$, but there is also a separate 'blob' that 
moves more to the Southwest with a projected speed of 450 km s$^{-1}$, 
see Fig.~\ref{fig4}. The TRACE image in Fig.~\ref{fig4} is at 
03:02:04 UT, very near the time when the first fragmented radio 
type II burst band appeared.  

\begin{figure}
   \centering
   \includegraphics[width=8cm]{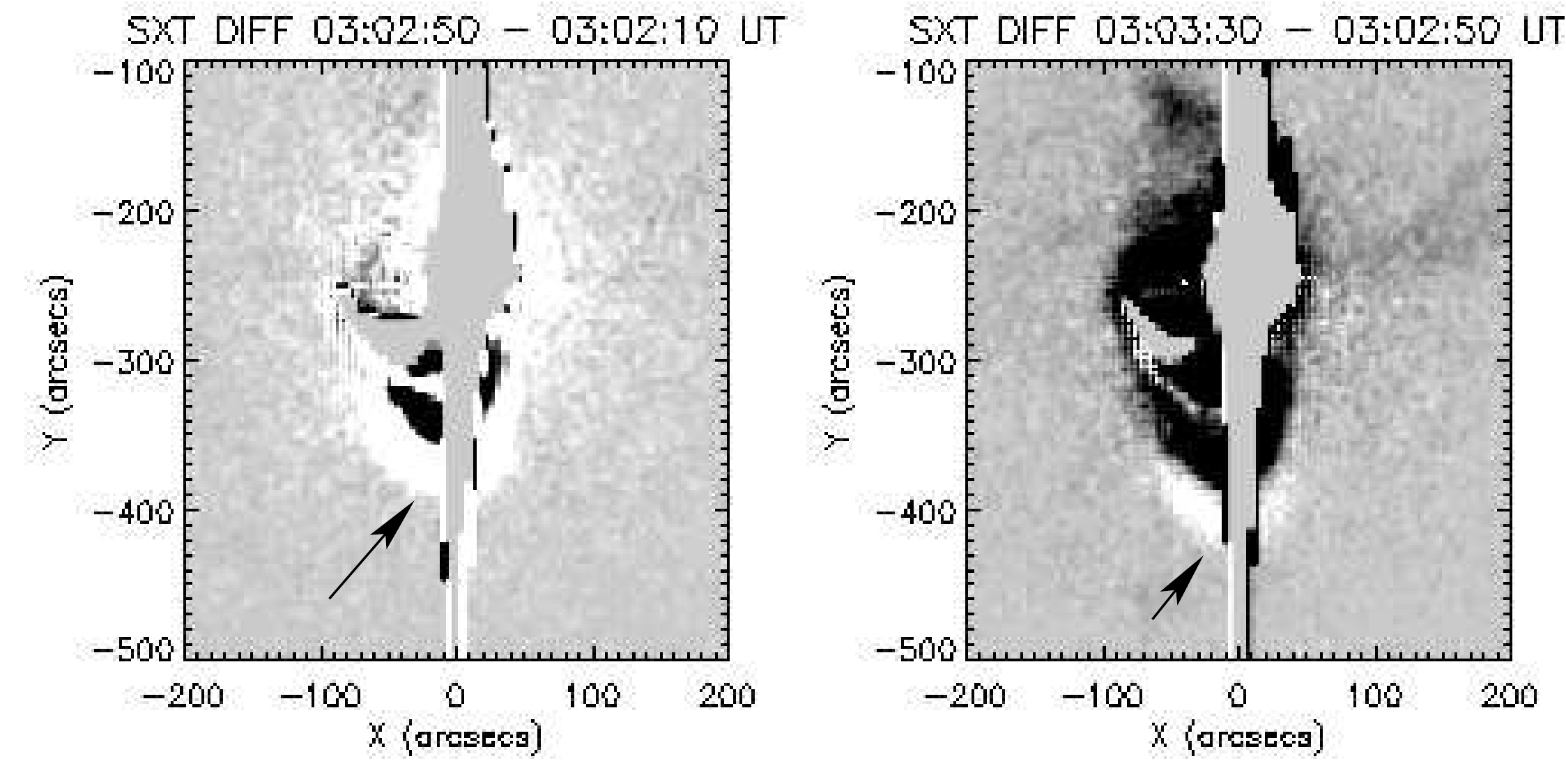}
      \caption{SXT difference images (AlMg filters) at 03:02:50 UT
              and 03:03:30 UT, showing the loop-like eruption front 
              in soft X-rays. The later difference image is
              near the time of the appearance of the 'regular' metric
              type II burst at 03:03:35 UT. The projected 
              plane-of-the sky speed of the soft X-ray front is 
              about 650 km s$^{-1}$ (arrows point to the loop front).
              }
         \label{fig5}
   \end{figure}

Yohkoh SXT  observed the flare until 03:04 UT, when satellite night 
set in. The partial SXT frames between 03:02 and 03:03 UT reveal 
a loop-like front moving Southward at a projected speed of about 650 
km s$^{-1}$, see Figs.~\ref{fig4} and \ref{fig5}. 
The projected distances for the SXT loop-like front and the 
EUV blob, calculated from the eruption center where loop footpoints 
were visible prior to the event, are shown in Fig.~\ref{fig1}, together 
with the type II burst height estimates (at 10$\times$Saito atmospheric 
densities). The height of the first type II burst fragment is low 
compared to the SXT and EUV structure heights, but there is also some
uncertainty in the true density gradient. The fragment heights, if true, 
suggest that the shock may have travelled through the EUV and/or soft
X-ray structures. On the other hand, a low type II start height also 
suggests that the type II burst may have been emitted at the low-lying 
flanks of the shock. If, however, the speed of the propagating shock
remained in the range of 1\,000\,--\,850 km s$^{-1}$ during the whole
event, the fragmented burst could have started at heliocentric 
height of 1.1 R$_{\odot}$ ($\approx$\,70\,000 km above the photosphere),
which is just above the SXT loop-like front.    
   
As Yohkoh SXT was observing with two different filters, Be and AlMg, 
it is possible to estimate the emission measure $EM$ using the filter 
ratio, see, e.g., \citet{mctiernan93}.
The emission measure for the outermost SXT loop-like structure at 
03:02:18 UT is 6 $\times$ 10$^{43}$\,--\,10$^{44}$ cm$^{-3}$ per pixel. 
As one pixel is 2.5\arcsec and the loop width is approximately 
10\arcsec in the images, we can estimate the loop depth along the 
line of sight to be equal to the loop width, $\approx$\,4 pixels, 
and thus get an emitting volume $V$ of 2.3 $\times$ 10$^{25}$ cm$^{3}$ 
per pixel. Assuming an isothermal homogeneous source, the 
electron density $N$ can be calculated from $N \approx \sqrt{EM/V}$, 
which gives a value of about  1.6\,--\,2.1 $\times$ 10$^9$ cm$^{-3}$. 
For comparison, the fragmented radio emission near 03:02 UT indicated
electron densities of 1\,--\,2 $\times$ 10$^9$ cm$^{-3}$, a value 
similar to our soft X-ray loop densities.   
  
The SXT difference image at 03:03:30\,--\,03:02:50 UT (Fig.~\ref{fig5}) 
shows that the (projected) soft X-ray loop-like front was located  
approximately 180\arcsec \ from the center of the flaring region. 
This corresponds roughly to a distance of 126\,000 km. For comparison, 
the estimated height of the later, regular part of the type II burst 
source at 03:03:35 UT is about 152\,000 km. Even taking into account 
projection effects, the radio burst source is located higher 
than the soft X-ray loop-like front at that time.

%

\section{Coronal conditions}

If we assume a structure moving through a system of dense loops 
at a speed of 850 km s$^{-1}$ (similar to the inferred metric type II shock 
speed; see also Sect. 5), shock-acceleration could be created if the speed 
of the disturbance exceeded the local magnetosonic speed. In the solar corona, 
the local Alfv\'en speed $v_A$ is a good approximation for the magnetosonic 
speed, 
\begin{equation}
v_A \approx \frac{2 \times 10^{11} B}{\sqrt{n_e}} , 
\end{equation}
where $B$ is the magnetic field strength and $n_e$ is the local electron 
density, in gaussian units.
In order to make the disturbance velocity super-Alfv\'enic at the density
of the first type II fragment (2\,-\,3 $\times$ 10$^9$ cm$^{-3}$), the
magnetic field strength should be less than 22 G. At the time
of the third fragment, the electron density was 2\,-¸6 $\times$ 10$^8$ 
cm$^{-3}$, implying a field strength less than 10 G. 
The empirical scaling law presented by \citet{dulk78} gives the strength 
of the coronal magnetic field above active regions as 
$B$ = 0.5\,$\times$\,$h^{-1.5}$ G, where $h$ is the height above the 
photosphere expressed in solar radiae. Field strengths of 22 G would then 
appear at about heights of 56\,000 km and 10 G at about 95\,000 km.
The outermost SXT loop heights, approximated from the observed projected 
distances between the loop footpoints and the loop fronts, were about 
66\,000 km at the time of the first type II fragment and about 100\,000 km 
at the time of the third fragment. The EUV blob was located lower,
at about 47\,000 and 71\,000 km, respectively. The estimated field strengths 
imply that the shock could have been located at about the SXT loop heights 
at the time of the fragmented type II burst. 

Basically, any propagating shock would be able to excite a type
II burst, since the type II characteristics mainly reflect the medium
where the shock is propagating. 
The SXT loop-like front could have been a low-coronal signature 
of the CME \citep{rust76,vrsnak04b,temmer08}, which drives the shock.  
The shock could also have been ignited by the flare, and propagated
as a blast wave. In this scenario, if the SXT loop-like front 
roughly corresponds to the CME front, the type II fragmented emission
would be excited at segments where the shock overtakes some part of
the CME -- and the type II fragments would be attributed to the 
inhomogeneities within the erupting structure.
One difference between shock types is that a blast wave should weaken 
in time as it does not gain energy on the way, and because it also  
expands the energy is divided into a larger area. 
Eventually a blast wave shock will become so weak that it cannot 
accelerate electrons any more, leading to a slow fade out of the 
type II burst. 

The type II burst disappearance from the spectrum was quite sudden 
at 03:05:15 UT. The abrupt end suggests that the shock may not
have been weakening but instead it entered a region where the magnetic
field orientation was very different, e.g., there was a transit
from a perpendicular to a parallel shock region. Shock-drift
acceleration is very efficient for quasi-perpendicular shocks,
and a change in the geometry could stop accelerating particles. 
Of course, this could happen both to a driven shock and a 
freely propagating blast wave shock (before it weakens).      
In the case where a blast wave shock overtakes a CME, type II
emission could end after the overtake \citep{gary84}.

\section{MHD modeling}

We utilize numerical simulations to study the shock structure induced
by an erupting CME in a model corona including dense loops. Since we
are interested in modeling the first minutes of the eruption, we
consider a local model of the low corona. The time-dependent, ideal
MHD equations augmented with gravity are solved in a
two-dimensional Cartesian domain.  The details of the model have been
presented by \citet{Pomoell2008}, who employed a largely similar model
to study the waves excited by an erupting flux rope.

The magnetic field configuration consists of a flux rope and an
arcade-like background field. The field of the detached flux rope is
created by a line current of radius $r_0$ situated at distance $h$
above the solar surface, and the arcade structure of the background
structure follows the coronal arcade model of \citet{Oliver1993}. While
the density of the ambient corona decreases exponentially due to the
choice of constant gravitational acceleration, 
$g_\odot = 2.74 \times 10^4$ cm s$^{-2}$, and constant initial temperature 
$T_0$ = 1.5 $\times$ 10$^6$ K, the density inside the flux rope is 
enhanced to model denser filament material.

To include coronal loops with an increased density into the model, we
apply a simple approach in which the magnetic field strength is
decreased in the denser loops so that the total pressure is constant
at a constant height.  For the details of this procedure, we refer to
the work of \citet{Odstrcil2000}.  Note that in real coronal loops,
both the magnetic field strength and density are enhanced compared
with the ambient corona. We discuss in section 6 the implications of
this when comparing the results of the simulations with the
observations.

\begin{table}
\caption{Alfv\'en speed in km/s at four locations $(x,y)$ 
for the two initial conditions}
\label{alfvenspeedtable} 
\centering 
\begin{tabular}{c c c c c} 
\hline\hline 
Run & $(-0.16,1)$ & $(0,1.1)$ & $(0,1.2)$ & $(0,1.45)$\\ 
\hline 
A      & 335 & 515 & 220 & 571\\ 
B \& C & 457 & 703 & 30  & 779\\
\hline
\end{tabular}
\end{table}

\begin{figure}[b] 
   \centerline{\bf \hspace{0.09\textwidth}  \color{black}{Density} 
   \hspace{0.07\textwidth}  \color{black}{Alfv\'en speed}
   \hspace{0.05\textwidth}  \color{black}{$\rho$} 
   \hspace{0.044\textwidth}  \color{black}{$\nu_A$}   \hfill }
   \centerline{\includegraphics[width=8cm]{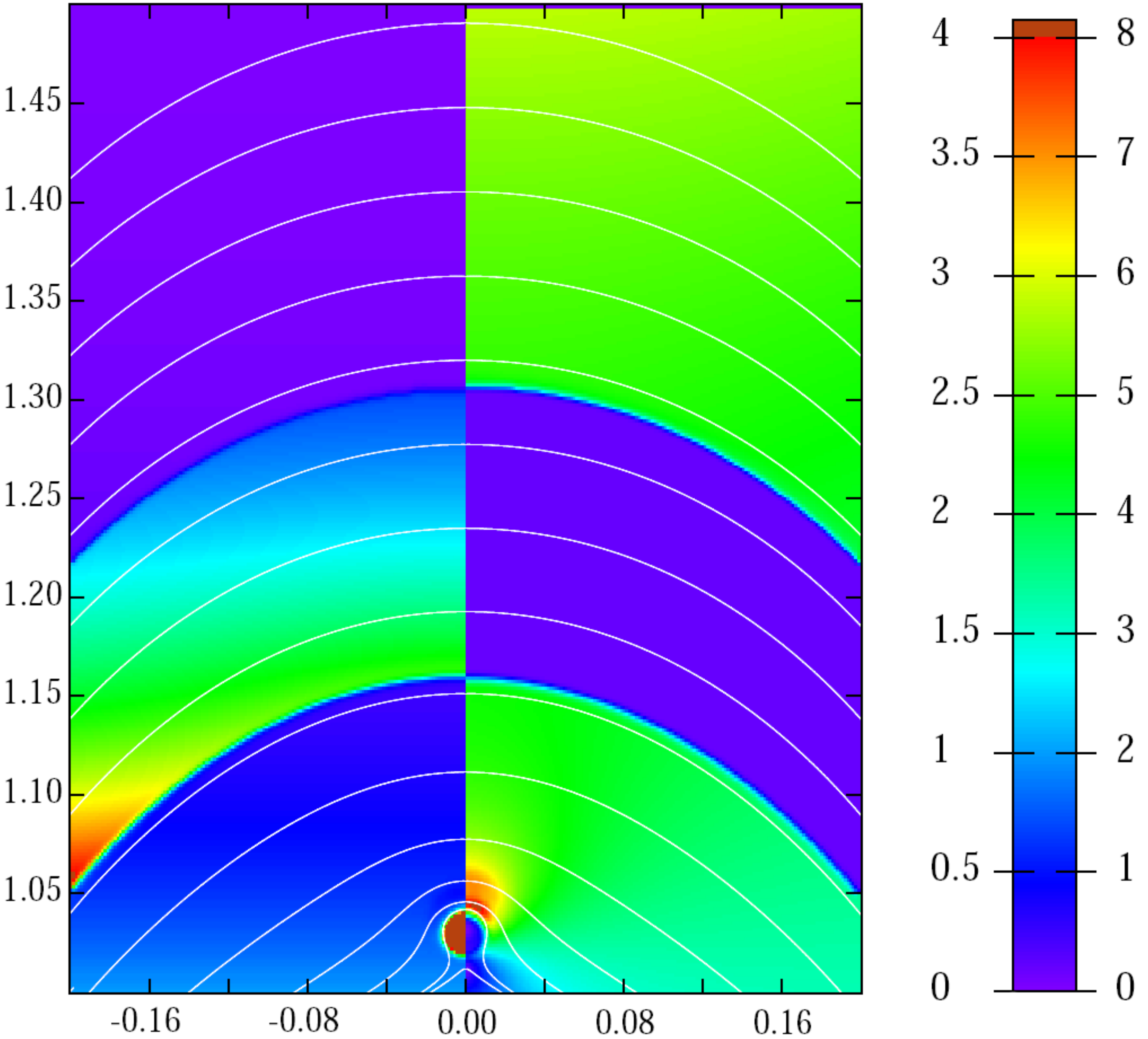}}
   \vspace{-0.22\textwidth}
   \centerline{\hspace{0.0\textwidth}  \color{black}{$y$} \hfill }
   \vspace{0.22\textwidth}
   \vspace{-0.022\textwidth}
   \centerline{ \hspace{0.19\textwidth}  \color{black}{$x$} \hfill }
   \vspace{-0.01\textwidth}
   \caption{The initial state of the simulation: density (left half) and
            Alfv\'en speed (right half). The white lines depict magnetic
            field lines. The units in the color bar are for the density 
            (left values in the color bar)
            $1.67 \times 10^{-15} \mbox{ g\,cm}^{-3}$ and for the Alfv\'en 
            speed (right values in the color bar)
            $143 \mbox{ km\,s}^{-1}$, the isothermal sound speed. 
            Note the clipping of the color bar; a brown 
            color indicates values larger than the color bar maximum.}
   \label{initial-density}
\end{figure}

Fig. \ref{initial-density} shows the initial distribution of the
density and Alfv\'en speed $\nu_A$ (isothermal sound speed) 
with some field lines of the magnetic field superimposed.  Table
\ref{alfvenspeedtable} gives the Alfv\'en speed at four locations for
the two initial conditions considered (see section 5). Note that
$\nu_A$ increases as a function of height (away from the fluxrope),
but varies less inside the loop.

\begin{figure}[!ht]   
 \centerline{\bf \hspace{0.12\textwidth}  \color{black}{Density} 
 \hspace{0.07\textwidth}  \color{black}{Speed}
 \hspace{0.09\textwidth}  \color{black}{$\rho$,$\nu$}\hfill }
 \centerline{\includegraphics[width=0.345\textwidth,clip=]{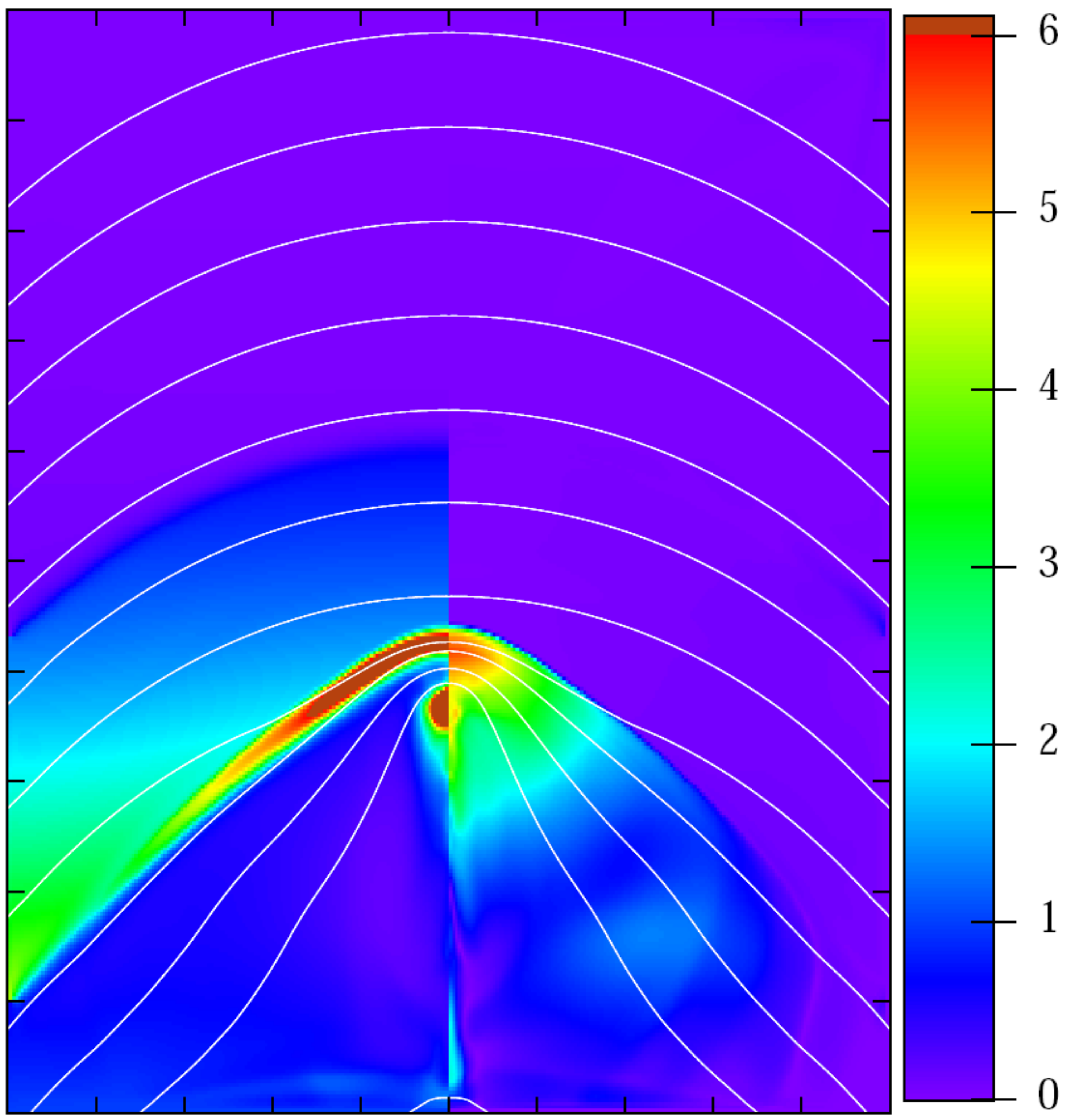}}
 \centerline{\hspace{-0.002\textwidth}\includegraphics[width=0.343\textwidth,clip=]{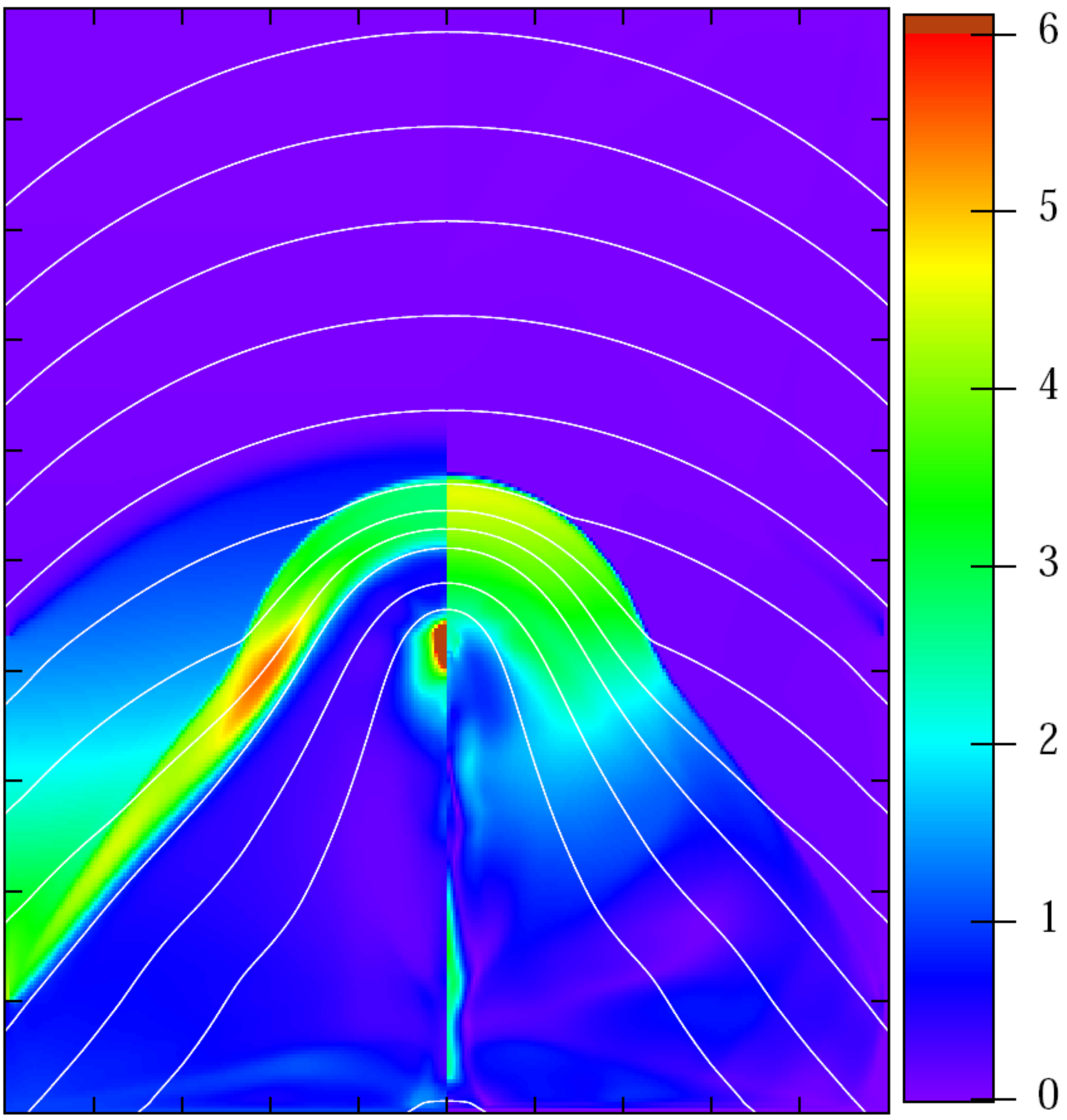}}
 \centerline{\includegraphics[width=0.345\textwidth,clip=]{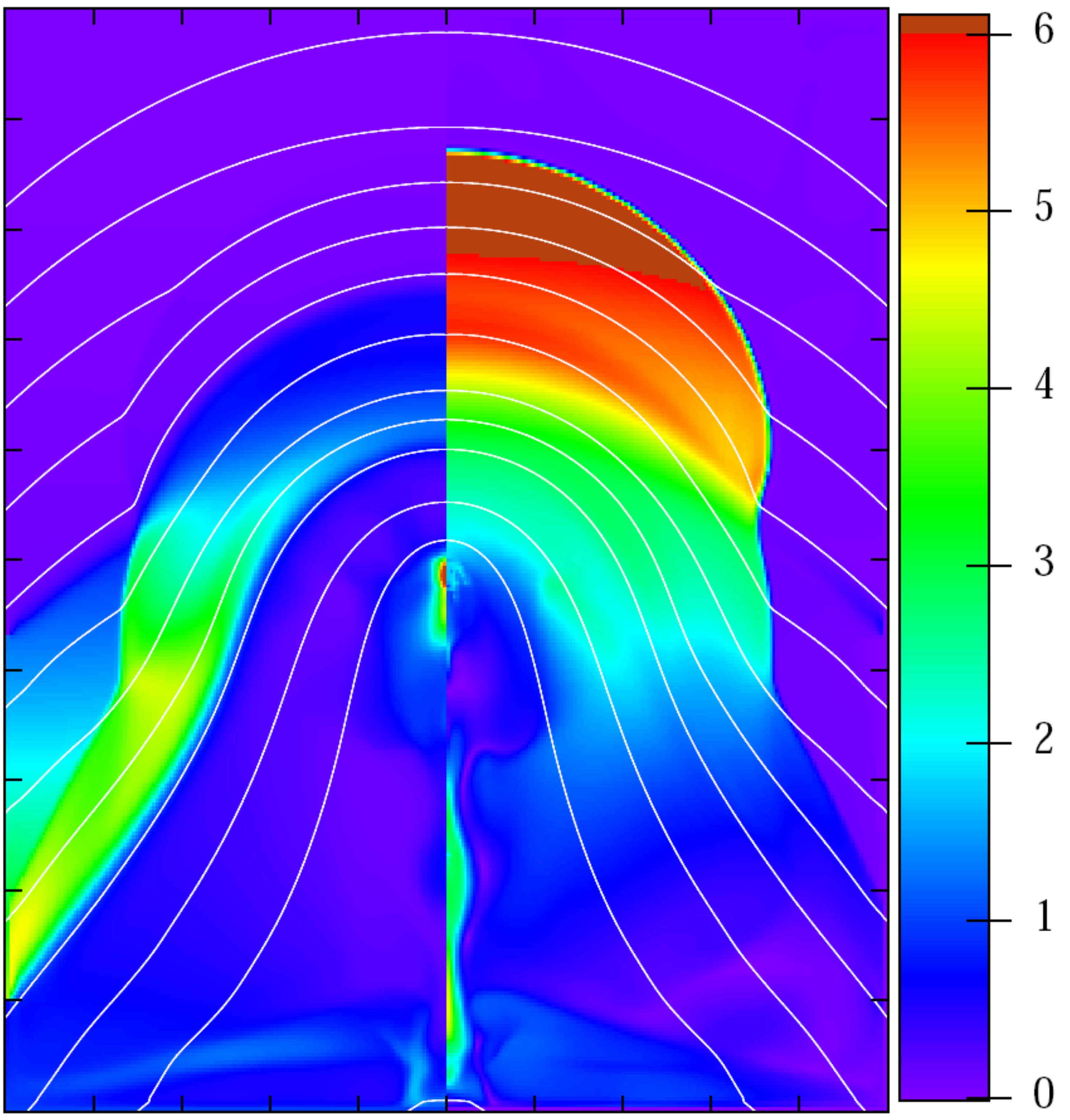}}
 \vspace{-1.045\textwidth}
 \centerline{\Large \bf \hspace{0.01\textwidth}  \color{black}{(a)} \hfill }
 \vspace{1.045\textwidth}
 \vspace{-0.708\textwidth}
 \centerline{\Large \bf \hspace{0.01\textwidth}  \color{black}{(b)} \hfill }
 \vspace{0.708\textwidth}
 \vspace{-0.37\textwidth}
 \centerline{\Large \bf \hspace{0.01\textwidth}  \color{black}{(c)} \hfill }
 \vspace{0.292\textwidth}
 \caption{ The density (left half) and speed (right half) at three 
 different times, $t=220$ (a), $t=270$ (b) and $t=330$ (c) seconds. 
 The white lines depict the magnetic field lines. The units in the color 
 bar are for the density $1.67 \times 10^{-15} \mbox{ g\,cm}^{-3}$ and for 
 the speed $143 \mbox{ km\,s}^{-1}$. Note that both quantities are plotted 
 using the same color bar.}
 \label{snapshots}
 \end{figure}

The flux rope is made to erupt by invoking an artificial force that
acts on the flux rope plasma elements during the simulation, similar
to the study of \citet{Pomoell2008}.  The Cartesian grid of the
simulation consists of $300 \times 300$ cells in the $x\times y=[$-0.2
  R$_\odot$, 0.2 R$_\odot] \times [$1.0 R$_{\odot}$, 1.5 R$_\odot]$
simulation domain. The boundary values are kept fixed during the
simulation.

\begin{figure*}[t]  
\includegraphics[width=8.5cm]{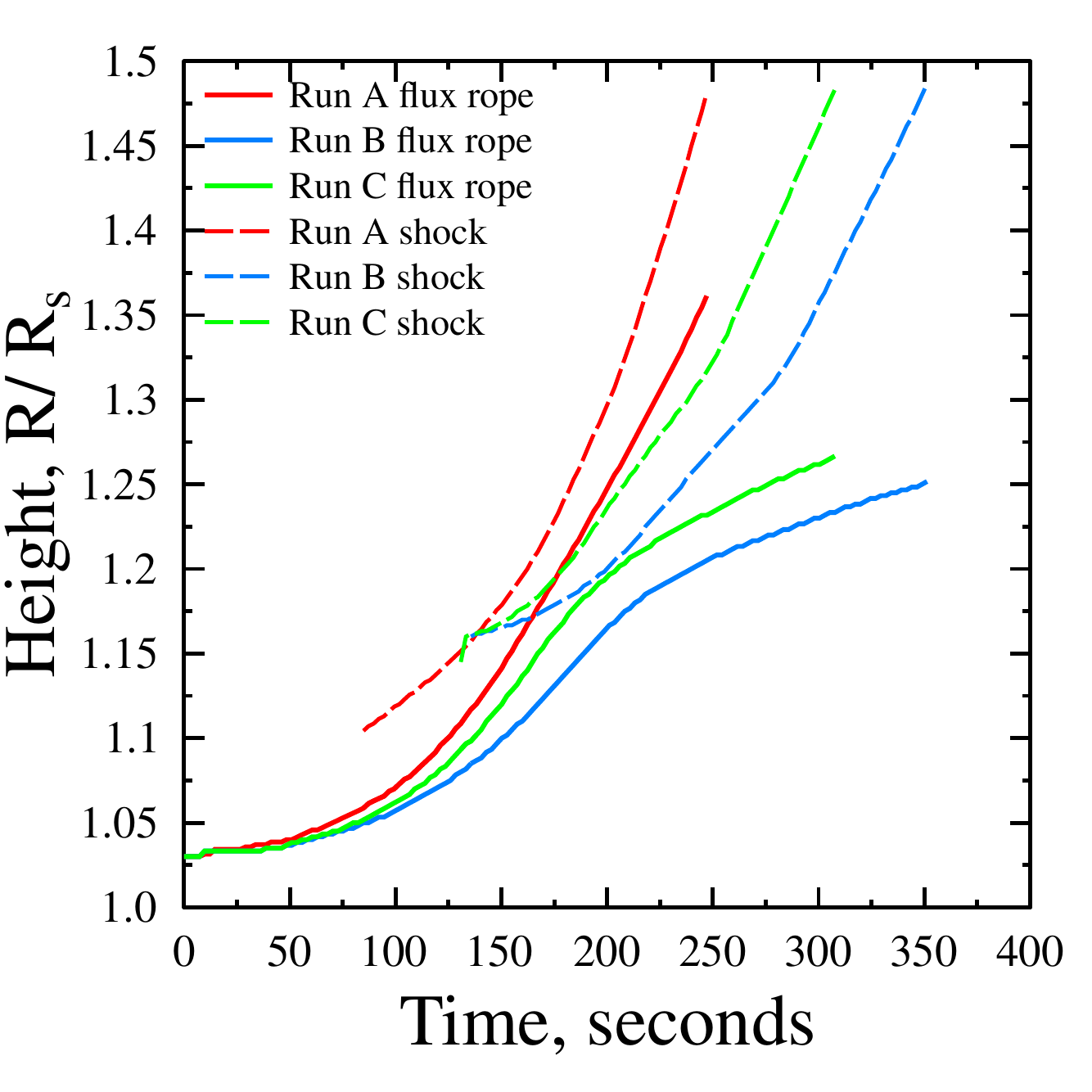}
\includegraphics[width=8.5cm]{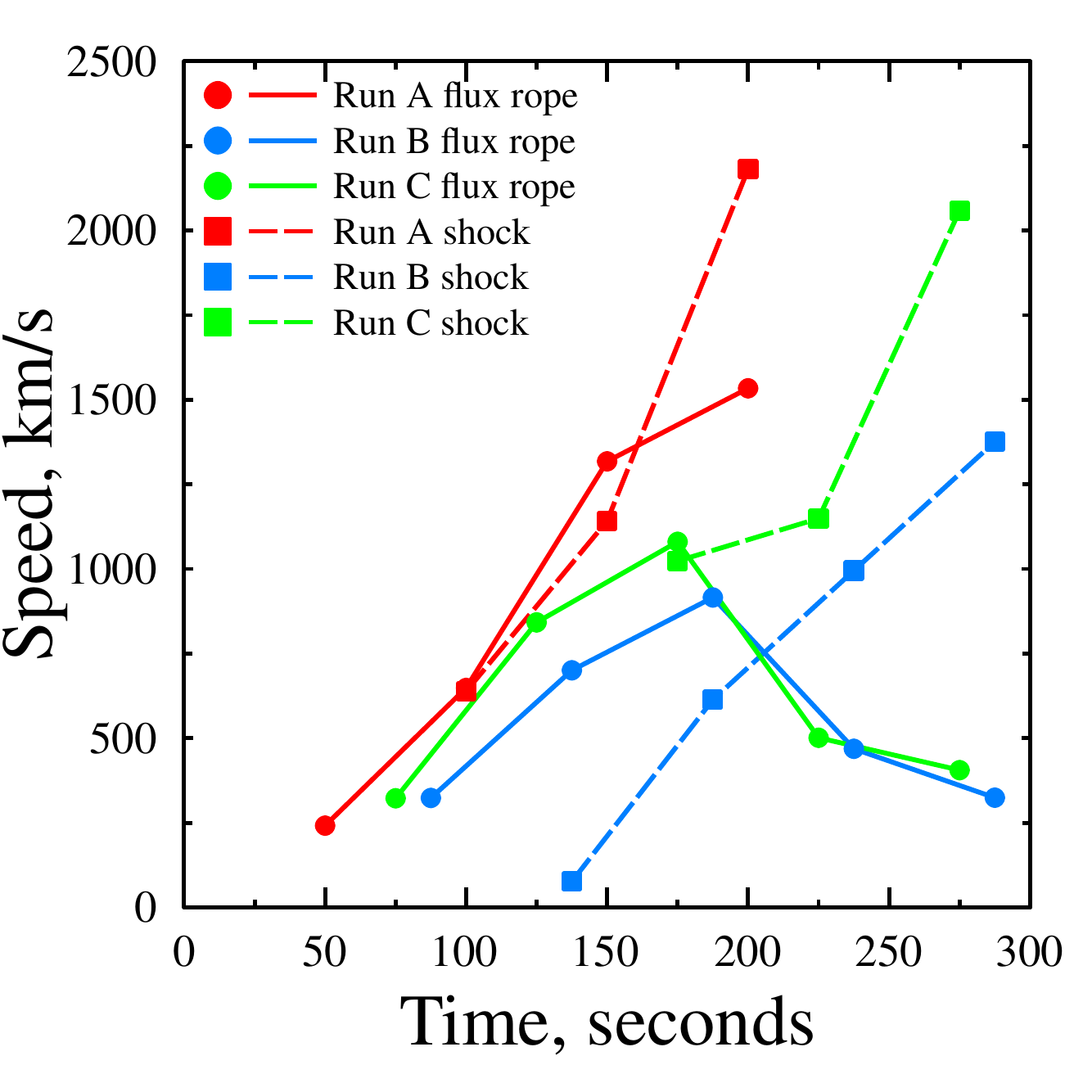}
   \caption{Left: Plot of the height-time data of the flux rope
            (solid line) and shock front (broken line) for the three
            simulation runs.  Red corresponds to run A, green to run B
            and blue to run C. 
            Right: Plot of the corresponding flux rope and shock 
            speed. }           
   \label{height-time-plot}
\end{figure*}

\section{Results  of the simulations}

Fig. \ref{snapshots} shows the mass density and speed at three
different times, $t=220, 270$ and $330$ seconds.  The dynamics of the
eruption is as follows: As the flux rope starts to rise, a
perturbation is formed around the flux rope. 
Due to the gradient in the Alfv\'en speed and the increasing
speed of the flux rope, the wave steepens to a shock ahead of the
flux rope.  However, the strength of the shock remains weak in the
area below the loop. When the shock reaches the dense loop, it
strengthens and slows down quickly due to the low Alfv\'en speed in
the loop (panel a). The erupting filament continues to push the loop
structure ahead of it, acting as the driver of the shock. 
Thus, the speed of the shock is roughly that of the displaced loop 
structure when propagating in the region of low $\nu_A$. As the filament
decelerates, the displaced loop and shock escape from the filament
(panel b). When reaching the region of higher $\nu_A$, the speed of
the shock increases with the increasing $\nu_A$, and escapes from
the propagating loop structure (panel c).  

To quantify the description above, we plot in
Fig.~\ref{height-time-plot} the height-time profile of the filament
and the shock. We consider three different runs with slightly
different parameters. Run B (blue color coding) and C (green color
coding) are identical but for the magnitude of the artificial
accelerating force, which is larger for run C. Run A has the same
magnitude of the accelerating force as run C, but the density contrast
between the loop and ambient corona as well as the strength of the
magnetic field is lower for run A. 
The speeds of the flux rope and shock are also shown in
Fig.~\ref{height-time-plot}. 

In all three cases, the shock is strong while it propagates in the
loop, with the compression ratio remaining between 3\,--\,3.5.  However,
the behaviour after the shock exits the loop depends on the simulation
run. For run A, the shock only slightly decreases in strength, while
for runs B and C the compression ratio quickly decreases to
approximately 2.2. Interestingly, the shock strength remains at this
value until the end of the simulation.

\begin{figure}[t]  
\centerline{\includegraphics[width=8cm]{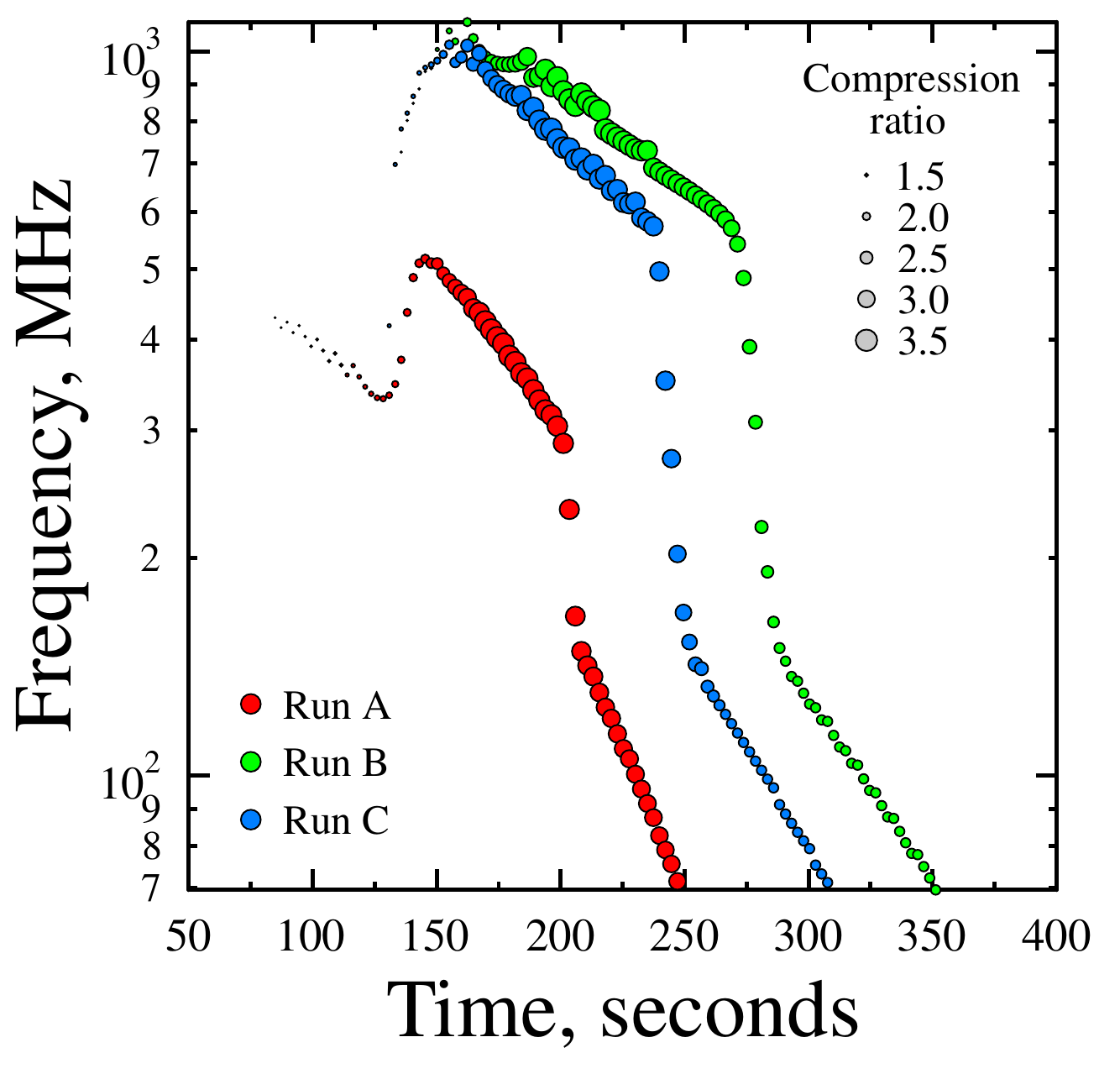}}
\caption{Radio track of the three simulation runs. The size of the 
 marker indicates the compression ratio of the shock.}
\label{radio-oneloop-plot}
\end{figure}

In Fig.~\ref{radio-oneloop-plot}, we plot the radio emission produced
by the shock assuming that the emission is produced immediately in
front of the leading edge shock. The emission lane shows a frequency
drop in accordance with the exponentially decreasing
density. Additionally, we indicate the compression ratio of the shock
by the size of the marker.

In addition to performing simulations with one dense loop, we also
considered simulations with two loops. Adding another dense loop above
the existing loop does not alter the overall dynamics of the
eruption. However, the shock structure becomes very complex as the
shock enters the second loop. We refrain from showing the results, but
note that a similar plot as Fig. \ref{radio-oneloop-plot} in the
two-loop case shows two distinct tracks, corresponding to the shock
propagating in the dense loops. In between the loops the frequency
drops significantly.

\section{Discussion and conclusions}

Observations on 13 May 2001 show that:
    \begin{itemize}
      \item The speed of the white-light, plane-of-the-sky CME 
            (\mbox{430 km s$^{-1}$}) was similar to the projected speed of the 
            erupting filament  (380 km s$^{-1}$) and the EUV ``blob''
            (450 km s$^{-1}$). 
            The CME most probably consisted of the ejected filament, and 
            the filament propagated in the wake of the soft X-ray loop-like  
            front
      \item The projected speed of the soft X-ray loop-like front 
            (\mbox{650 km s$^{-1}$}) and the estimated speed of the 
            type II burst source ($>$\,850 km s$^{-1}$) were higher 
            than the CME and EUV structure speeds   
      \item The estimated soft X-ray loop densities agree with the plasma 
            densities of the fragmented metric type II burst. 
            Also magnetic field strengths are in agreement with a 
            super-Alfv\'enic shock at speed $\approx$\,850 km s$^{-1}$ 
            near the soft X-ray loop heights  
      \item The fragmented type II burst changed appearance to
            a ``regular'' type II burst at a time when the propagating 
            soft X-ray loop-like front approached the outer boundary of the 
            active region
      \item The type II burst disappeared abruptly, without any
            weakening of the emission.
   \end{itemize}

\noindent{The MHD simulations that we have performed for a flux rope
  accelerated in a corona with dense loops show that}:
\begin{itemize}
\item The CME eruption drives a shock wave ahead of it through the corona 
\item The shock wave propagates faster through the corona than its driver
\item In the dense loops, the shock becomes significantly stronger
\item Under right conditions, the shock strength decreases rapidly as
  the shock exits the loop.
\end{itemize}

It is consistent with the simulation model that the high speed of the
loop-like soft X-ray front actually corresponds to a strong shock wave
propagating through the loop. When outside the active region, the
changing geometry no longer favours shock acceleration and the type II
burst dies out. Together with the assumption that the shock wave
generates conditions that produce type II bursts, our MHD model
produces burst lanes that correspond qualitatively to those observed
in the event. However, in order for the radio signature to become
fragmented as is observed, the conditions for plasma emission have to
be somehow more favourable inside the loop than in the interloop
area. The obvious assumption, consistent with our simple simulation
model, is that the shock strength decreases significantly in the space
between the denser loops. This requires specific conditions, such as a
very large difference in the Alfv\'en speed between the loop and
interloop area, or a rapidly increasing Alfv\'en speed in the
interloop area.

Our simplified two-dimensional model of coronal loops generates the
favourable conditions for shock formation as it necessarily yields
weak magnetic fields in the dense loops to achieve lateral pressure
balance. This is obviously too restrictive an assumption, as real
coronal loops involve both increased densities and magnetic
fields. Thus, the strength of the shock in our model does not
necessarily carry over to a more realistic model of the
corona. However, the overlying coronal structure may actually be quite
fragmented containing loops with different densities and magnetic
fields, and our simulation model indicates that those with the lowest
Alfv\'en speed produce the strongest shock. Note also that the
strength of the shock is not the only factor affecting the radio
emission from the shock. Other factors, such as the density of the
electrons in the loop, are important for the plasma emission mechanism
as well. Thus, the densest loops in the corona overlying the active
region seem most probable sites of plasma emission.

It is also possible and even plausible that the three-dimensional
structure of the corona plays a significant role. For instance, the
different fragments of the bursts could originate from different loops
that are not on top of each other. However, even in this case the
interloop area must have specific conditions but not necessarily
related to the strength of the shock, in order to cause the burst to
suddenly stop emitting.

In conclusion, using results from multi-wavelength data analysis and
numerical MHD simulations, we have established that the unusual metric
type II radio burst on 13 May 2001 can be explained by a model, where
a coronal shock driven by a mass ejection passes through a system of
dense loops overlying the active region, where the ejection
originates.

\begin{acknowledgements}
We thank the referee, B. Vr\u{s}nak, for valuable comments and
suggestions on how to improve the paper.  
We have used in this study radio observations obtained from the 
Hiraiso Solar Observatory (National Institute of Information and 
Communications Technology, Japan) and we thank the staff for making 
the radio spectra available at their web site. The LASCO CME Catalog 
is generated and maintained at the CDAW Data Center by NASA and the 
Catholic University of America in cooperation with the Naval Research 
Laboratory. We are grateful to the TRACE, Yohkoh, SOHO EIT and LASCO 
teams for making their data available. 
J.P. acknowledges financial support from the Vilho, Yrj\"o, and Kalle
 V\"ais\"al\"a foundation.
\end{acknowledgements}


\begin{thebibliography}{}
\bibitem[\protect\citeauthoryear{Bougeret et al.}{1995}]{bougeret95}
Bougeret, J.-L., Kaiser, M.L., Kellogg, P.J., et al. 1995, 
Space Sci. Rev., 71, 231
\bibitem[\protect\citeauthoryear{Brueckner et al.}{1995}]{brueckner95} 
Brueckner, G.E., Howard, R.A., Koomen, M.J., et al. 1995, 
Solar Phys., 162, 357
\bibitem[\protect\citeauthoryear{Cairns et al.}{2003}]{cairns03}
Cairns, I.H., Knock, S.A., Robinson, P.A., \& Kuncic, Z. 2003, 
Space Sci Rev., 107, 27
\bibitem[\protect\citeauthoryear{Cane \& Erickson}{2005}]{cane05}
Cane, H.V., \& Erickson, W.C. 2005, ApJ, 623, 1180
\bibitem[\protect\citeauthoryear{Cliver et al.}{1999}]{cliver99}
Cliver, E.W., Webb, D.F., \& Howard, R.A. 1999, Solar Phys., 187,89 
\bibitem[\protect\citeauthoryear{Dauphin et al.}{2006}]{dauphin06}
Dauphin, C., Vilmer, N., \& Krucker S. 2006, A\&A, 455, 339
\bibitem[\protect\citeauthoryear{Dulk \& McLean}{1978}]{dulk78}
Dulk, G.A., \& McLean, D.J. 1978, Solar Phys., 57, 279 
\bibitem[\protect\citeauthoryear{Gary et al.}{1984}]{gary84}
Gary, D.E., Dulk, G.A., House, L., et al. 1984, A\&A, 134, 222 
\bibitem[\protect\citeauthoryear{Handy et al.}{1999}]{handy}
Handy, B.N., Acton, L.W., Kankelborg, C.C., Wolfson, C.J., Akin, D.J., 
Bruner, M.E., et al. 1999, Solar Phys., 187, 229
\bibitem[\protect\citeauthoryear{Khan \& Aurass}{2002}]{khan02}
Khan, J.I. \& Aurass, H. 2002, A\&A, 383, 1018 
\bibitem[\protect\citeauthoryear{Klassen et al.}{2003}]{klassen03}
Klassen, A., Pohjolainen, S. \& Klein, K.-L. 2003, 
Solar Phys., 218, 197
\bibitem[\protect\citeauthoryear{Klein et al.}{1999}]{klein99}
Klein, K.-L., Khan, J.I., Vilmer, N., et al. 1999, 
A\&A, 346, L53
\bibitem[\protect\citeauthoryear{Kosugi et al.}{1991}]{kosugi91}
Kosugi, T., Masuda, S., Makishima, K., et al. 1991, Solar Phys., 136, 17
\bibitem[\protect\citeauthoryear{Lin et al.}{2006}]{lin06}
Lin, J., Mancuso, S., \& Vourlidas, A. 2006, ApJ, 649, 1110
\bibitem[\protect\citeauthoryear{Nelson \& Melrose}{1985}]{nelson85}
Nelson, G.J., \& Melrose, D.B., 1985, in Solar Radiophysics,
D.J. McLean and N.R. Labrum (eds.), Cambridge Univ. Press, 333
\bibitem[\protect\citeauthoryear{Mann \& Klassen}{2002}]{mann02}
Mann, G., \& Klassen, A. 2002, Proc. 10th European Solar
Phys. Meeting, ESA SP-506, 245 
\bibitem[\protect\citeauthoryear{Mann \& Klassen}{2005}]{mann05}
Mann, G., \& Klassen, A. 2005, A\&A, 441, 319
\bibitem[\protect\citeauthoryear{McTiernan et al.}{1993}]{mctiernan93}
McTiernan, J.M., Kane, S.R., Loran, J.M., Lemen, J.R., Acton, L.W., 
et al. 1993, ApJ, 416, L91 
\bibitem[\protect\citeauthoryear{Melrose}{1980}]{melrose80}
Melrose, D.B. 1980, Space Sci Rev., 26, 3
\bibitem[\protect\citeauthoryear{Odstr\u cil \& Karlick\'y}{2000}]
{Odstrcil2000} 
Odstr\u cil,~D., \& Karlick\'y,~M. 2000, A\&A, 359, 766 
\bibitem[\protect\citeauthoryear{Oliver et al.}{1993}]{Oliver1993} 
Oliver,~R., Ballester,~J.~L., Hood,~A.~W., Priest,~E.~R. 1993, 
A\&A, 273, 647
\bibitem[\protect\citeauthoryear{Pohjolainen}{2008}]{pohjolainen08}
Pohjolainen, S. 2008, A\&A, 483, 297 
\bibitem[\protect\citeauthoryear{Pohjolainen et al.}{2001}]{pohjolainen01}
Pohjolainen, S., Maia, D., Pick, M., et al. 2001, ApJ, 556, 421
\bibitem[\protect\citeauthoryear{Pohjolainen et al.}{2007}]{pohjolainen07}
Pohjolainen, S., van Driel-Gesztelyi, L., Culhane, J.L., 
Manoharan, P.K., \& Elliott, H.A. 2007, Solar Phys., 244, 167 
\bibitem[\protect\citeauthoryear{Pomoell et al.}{2008}]{Pomoell2008} 
Pomoell,~J., Vainio,~R., \& Kissmann,~R. 2008, Solar Phys., in press 
\bibitem[\protect\citeauthoryear{Rust \& Hildner}{1976}]{rust76}
Rust, D.M. \& Hildner, E. 1976, Solar Phys., 48, 381
\bibitem[\protect\citeauthoryear{Saito}{1970}]{saito70}
Saito, K. 1970, Ann. Tokyo Astr. Obs., 12, 53
\bibitem[\protect\citeauthoryear{Subramanian \& Ebenezer}{2006}]{subra06}
Subramanian, K.R. \& Ebenezer, E. 2006, A\&A, 451, 683
\bibitem[\protect\citeauthoryear{Temmer et al.}{2008}]{temmer08}
Temmer, M., Veronig, A.M., Vr\u{s}nak, B., et al. 2008, ApJ, 673, L95
\bibitem[\protect\citeauthoryear{Tsuneta et al.}{1991}]{tsuneta}
Tsuneta, S., Acton, L., Bruner, M., et al. 1991, Solar Phys., 136, 37
\bibitem[\protect\citeauthoryear{Vr\u{s}nak}{2005}]{vrsnak05}
Vr\u{s}nak, B. 2004, EOSTr 86, 112
\bibitem[\protect\citeauthoryear{Vr\u{s}nak et al.}{2002}]{vrsnak02}
Vr\u{s}nak, B., Magdaleni\u{c}, J., Aurass, H., \& Mann, G. 2002,
A\&A, 396, 673	
\bibitem[\protect\citeauthoryear{Vr\u{s}nak et al.}{2004a}]{vrsnak04a}
Vr\u{s}nak, B., Magdaleni\u{c}, J., \& Zlobec, P. 2004a, A\&A, 413, 753
\bibitem[\protect\citeauthoryear{Vr\u{s}nak et al.}{2004b}]{vrsnak04b}
Vr\u{s}nak, B.; Mari\u{c}i\'c, D., Stanger, A.L., \& Veronig, A.
2004b, Solar Phys., 225, 355
\bibitem[\protect\citeauthoryear{Warmuth}{2007}]{warmuth07}
Warmuth, A. 2007, in The High Energy Solar Corona: Waves, Eruptions, 
Particles, ed. K.-L. Klein, A.L. MacKinnon, Lect. Notes Phys. 725,
Springer, 107
\end{thebibliography}
\end{document}